\documentclass[twocolumn,aps,pra,floatfix,showpacs,noshowkeys,epsfig,graphics,natbib]{revtex4}%
\usepackage{graphicx}
\usepackage{amsmath}
\usepackage{amsfonts}
\usepackage{amssymb}

\newcommand{\newc}{\newcommand}
\newc{\beq}    {\begin{equation}}
\newc{\eeq}    {\end{equation}}

\newc{\beqa}    {\begin{eqnarray}}
\newc{\eeqa}    {\end{eqnarray}}
\newc{\bs}    {\section}
\newc{\no}    {\\ \nonumber}

\newc{\st}    {\stackrel}

\begin{document}

\title{N=2 PNGB Quintessence Dark Energy}

\author{Keunsu Cheon}
\email{kscheon@daejin.ac.kr}
\affiliation{Division of Mathematics and Physics, Daejin University, Pocheon, Gyeonggi 11159, Korea}

\author{Jungjai Lee}
\email{jjlee@daejin.ac.kr}
\affiliation{Division of Mathematics and Physics, Daejin University, Pocheon, Gyeonggi 11159, Korea}
\affiliation{Korea Institute for Advanced Study 85 Hoegiro, Dongdaemun-Gu, Seoul 02455, Korea}

\date{\today}
\begin{abstract}
 
  In this paper we show that a pseudo-Nambu-Goldstone boson (PNGB) quintessence of spontaneous symmetry breaking (SSB) is responsible for an epoch of the late time cosmic acceleration. We suggest that an N=2 PNGB quintessence with SSB can give rise to the fast-roll cosmic acceleration in evolution of the Universe. In the N=2 PNGB quintessence model, the standard slow-roll condition in which the absolute value of the tachyonic mass squared $|m_q^2|$ of this model is much less than the square of the Hubble constant $H^2$, is broken down to the fast-roll condition, $|m_q^2|=O(H^2)$. However, it is shown that the fast-rolling associated with the N=2 PNGB quintessence of SSB can be led to the epoch of the late time cosmic acceleration. Indeed, in our PNGB model, this epoch can be quite long lasting since the mass of the N=2 PNGB quintessence field is extremely small as $m_q \approx 10^{-33} \rm{eV}$.
\end{abstract}

\maketitle

\section{Introduction}
The dark energy problem is one of the most important hard problems to solve in elementary particle physics and modern cosmology \cite{Peebles:1999rt}.
The observations related to Type Ia supernova \cite{Riess:1998rt,Perlmuter:1999rt}, cosmic microwave background (CMB) \cite{Spergel:2007rt} and large scale structure (LSS) \cite{Tegmark:2006rt} all support that the current Universe is experiencing cosmic acceleration, which is driven by a dark energy  with negative pressure.

One interpretation of the dark energy is the cosmological constant, which is determined by an energy scale for example the reduced Planck scale $M_{p}=2.43\times10^{18}\rm{GeV}$ in which  vacuum energy density is given as $M^4_{p}$. However, the cosmological constant has its own problem. It suffers from a severe fine-tuning problem \cite{Weinberg:1999rt,Carroll:1999rt}. This needs an extreme fine-tuning which adjusts vacuum energy density by canceling the zero-point energy of the vacuum to very high precision in order to give the presently observed dark energy density $\rho_{ob}\simeq(3\rm{meV})^4$. 

Another possibility of the dark energy is that a pseudo-Nambu-Goldstone boson (PNGB) can act as dark energy, which is currently dominating the energy density of the Universe and subsequently relaxing to its ground state \cite{Frieman:1995rt}. As usual the PNGB quintessence characterized by the domain wall number N is determined with the two dynamical conditions,
a spontaneous global U(1) symmetry breaking energy scale $f$ and an explicit symmetry breaking energy scale $M$ for example the scale of background instanton physics such as H instanton \cite{Ovrut:1991rt} or hidden sector gauge instanton \cite{Kim:2003rt}. 
The PNGB models have theoretical benefits as compared to the other quintessence models. In these models, the shift symmetry of the PNGB prevents quantum corrections due to the other fields from spoiling flatness of the effective potential and it from mediating a fifth force of gravitational strength \cite{Adak:2014rt}. So far many authors have studied the possibility of the PNGB quintessence providing the dark energy \cite{Hall:2000rt,Barbieri:2000rt,Hung:2000rt,Abrahamsek:2008rt}. Kim and Nilles have proposed the quintessential axion as the candidate of dark energy, which is composed of a linear combination of two axions through the hidden sector supergravity breakdown \cite{Kim:2003rt,Copeland:2006rt}.

The PNGB models however have difficulties of their own. Usually particle physics models such as the PNGB models \cite{Frieman:1995rt,Choi:2000rt,Kim:2003rt,Kaloper:2006rt}, supergravity inspired models \cite{Nunes:2000rt,Aghababaie:2003rt,Fre:2002rt} and string axiverse model \cite{kamionkowski:2014rt} have been based on the assumption that a quintessence field driving the current cosmic acceleration satisfies the slow-roll condition $|m_q^2|\ll H^2$.  In Ref. \cite{Kollosh:2001rt} authors however found that quintessence models based on $\mathcal{N}=8$, $\mathcal{N}=4$ and $\mathcal{N}=2$ gauged supergravity violate the slow-roll condition $|m_q^2|\ll H^2$, and Linde soon proposed a new idea of a fast-roll inflation \cite{Linde:2001rt};
in the PNGB models and supergravity inspired models, the standard slow-roll condition $|m_q^2|\ll H^2$  is broken down to the fast-roll condition $|m_q^2|=O(H^2)$. In Ref. \cite{Frieman:1995rt} it is turned out that the PNGB acts like non-relativistic dark matter rather than dark energy in the near future. Consequently, in the quintessence-dominated era, the late time cosmic acceleration can be accomplished only when the initial conditions are fine-tuned for 
the PNGB field to slowly roll down near the top of its effective potential \cite{Kolda:1999rt,Svrcek:2006rt,Choi:2000rt,kamionkowski:2014rt}.

In this paper we show that an N=2 PNGB model of
spontaneous symmetry breaking (SSB) slightly could alleviate the fine-tuning problem of
the initial condition of the N=1 PNGB
quintessence models. It is also shown that the N=2 PNGB quintessence model can give the fast-roll cosmic acceleration in evolution of the Universe. Indeed, in the N=2 PNGB model, the late time cosmic acceleration can be sustained for a sufficiently long time since the mass of the N=2 PNGB quintessence field is extremely small as $m_q \approx 10^{-33} \rm{eV}$. We propose an idea that dark energy is originated from the N=2 PNGB quintessence of SSB. This idea linking dark energy to SSB is quite natural in the context of particle physics because SSB is a main concept when it comes to describing realistic theories of elementary particles.

Before turning to the details of the issue, let us briefly explain how SSB can work in the N=2 PNGB model, whereas it can not happen in the N=1 PNGB model. A spontaneous symmetry breaking is the way of obtaining an asymmetric ground state by arbitrarily selecting one of the degenerate vacuum states \cite{Mandl:2006rt}. In the N=1 PNGB model, its effective potential $V(Q)$ has only one minimum at $Q=0$, and thus spontaneous symmetry breaking can't happen (see FIG. 1). Classically the N=1 PNGB field (the gray ball down hill) starts to slowly roll down to the potential minimum, and subsequently converts to normal modes of oscillation around the potential minimum $Q=0$. In the context of quantum field theory, the N=1 PNGB field will produce massive real scalar particles with spin zero around the potential minimum $Q=0$.

The N=2 PNGB field (the gray ball on the central hill in FIG. 2) is not in that case. Here its effective potential $V(Q)$ has a local maximum at $Q=0$, and two local minima at $Q=\pm{\pi{F}}$, where $F$ is the axion decay constant with
$F = f/2$. We see that the ground state is not unique in the N=2 PNGB potential. If $Z_2$ symmetry $Q\leftrightarrow-Q$ is broken by the selection of one of the two degenerate vacuum states, the ground state of the system will be in one of two minima. In this case, the point is that initial field configuration to respect $Z_2$ symmetry might result from the
finite-temperature effects on the effective potential, as we shall see in the next section.
If so, classically, the N=2 PNGB field could settle down initially on the top of the effective
potential at $Q = 0$, and consequently taking field velocity
$\dot{Q}
= 0$. However, from the viewpoint of quantum field theory, the $Z_2$ symmetry of the N=2 PNGB field should be broken spontaneously because of growing quantum fluctuations of the field.

This paper is organized as follows. In section II, we summarize the dynamics of SSB for the N=2 PNGB field. In Section III, we show that the fast-roll cosmic acceleration in the N=2 PNGB quintessence of SSB can be responsible for the current cosmic acceleration. Section IV contains discussions and conclusions.
\begin{figure}[htbp]
\includegraphics[width=0.8\textwidth]{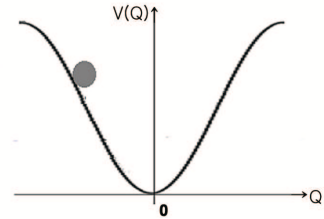}
\caption{N=1 PNGB potential, $V(Q)=M^4[1-\cos(\frac{Q}{f})]$: N=1 PNGB field (the gray ball down the hill) has only one minimum at $Q=0$. In this scheme, spontaneous symmetry break-
ing can’t happen.}
\label{rarfig}
\end{figure}

\begin{figure}[htbp]
\includegraphics[width=0.6\textwidth]{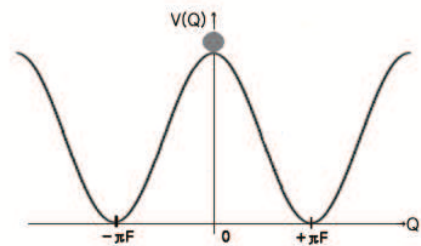}
\caption{N=2 PNGB potential, $V(Q)$=$M^4[1+\cos(2\frac{Q}{f})]$. N=2 PNGB field (the gray ball on the central hill) has one local maximum at $Q=0$, and  two local minima at $Q=\pm{\pi{F}}$.
}
\label{rasfig}
\end{figure}

\section{N=2 PNGB FIELD AND SPONTANEOUS SYMMETRY BREAKING}

For definiteness, we consider the N=2 PNGB model of SSB in the Minkowski spacetime. In the spacetime coordinates $x^{\mu}= (t, \vec{x})$ with the signature of metric $(-,+,+,+)$, the effective Lagrangian $\mathcal{L}$ for the PNGB field is given by
\beq
\mathcal{L}=-\frac{1}{2}\partial_{\mu}Q \partial^{\mu}Q - V(Q),
\label{LagranPo}
\eeq
where $\partial_{\mu}$ denotes the derivative with respect to $x^{\mu}$.
With the previously mentioned parameters $M$ and $f$, the PNGB effective potential $V(Q)$ in Eq. (\ref{LagranPo}) is given by
\beq
\label{PNGBpot}
V(Q)=M^4 \left[ 1+{\cos\left(2\frac{Q}{f}\right)}\right],~~f=\beta M_p.
\eeq
where $\beta$ is just a dimensionless parameter denoting the ratio of the global symmetry breaking scale $f$ to the reduced Planck mass scale $M_p$.
In this work it is assumed that the effective potential Eq. (\ref{PNGBpot}) is induced by a string instanton based on string axiverse \cite{Arva:2009rt,kamionkowski:2014rt}. The dynamical scale $M^4$ then is given by 
\beq
\label{Insta}
M^4=\mu^4e^{-S}.
\eeq
Here $\mu$ is $\mu=\sqrt{M_{SUGY}M_p}$, $M_{SUGY}$ the supersymmetry breaking scale, $S$ the string instanton action and $m_q$=$2(\frac{M^2}{f})$ mass of the N=2 PNGB field. In the
N=2 PNGB model, we take values as follows: $\beta =1$, $M=O(1\rm{meV})$. The N=2 PNGB potential $V(Q)$ then has two global minima at $Q=\pm\pi({f}/{2})$ and one local maximum at $Q=0$ with curvature $V''(0)=-m^2_q$.

Before turning to the details of the issue, let us briefly explore how initial field configuration to respect $Z_2$ symmetry could result from the
finite-temperature effects on the effective potential \cite{Gupta:1992rt}. we assume the existence of a
N=2 PNGB field with a Lagrangian
given as a function of temperature:
\beq
\mathcal{L}=-\frac{1}{2}\partial_{\mu}Q \partial^{\mu}Q - V(Q, T).
\label{LagranPo4}
\eeq
The finite-temperature effective potential $V(Q, T)$ in Eq. (\ref{LagranPo4}) is given as 
\beq
\label{PNGBpot3}
V(Q, T)=M^4 \left[1-c(T){\cos\left(2\frac{Q}{f}\right)}\right],
\eeq
where $c(T)$ is a function of temperature. We consider the case that $c(T)$ is given as follows:
\beq
\label{PNGBpot4}
c(T)=\left(\frac{T}{T_c}\right)^n-1,~n=2.
\eeq
In Eq. (\ref{PNGBpot4}) the critical temperature $T_c$ is
\beq
\label{PNGBpot5}
T_c=\gamma M,
\eeq
where $\gamma$ is order one parameter. At high temperatures $T\gg T_c\sim M$, $c(T)\simeq (T/T_c)^2$, and the effective mass squared of the N=2 PNGB field, $m_q^2=4\frac{M^4}{f^2}c(T)$ is positive. The finite-temperature effective potential $V(Q, T)$ then has a minimum at $Q=0$ and a maximum at $Q=\pi f/2$. If $c(T)$ is sufficiently large, the N=2 PNGB field mass $m_q$ is sufficiently large as well, and we expect that the N=2 PNGB field has rolled quickly down to the minimum of the effective potential at $Q=0$. When the temperature $T$ of the Universe is $T=T_c\sim M$, $c(T)=0$, the effective potential $V(Q, T)=M^4$, and the N=2 PNGB field is massless. At low temperatures $T\ll T_c\sim M$, $c(T)\simeq -1$, and the effective mass squared $m^2_q$ is negative. The finite-temperature effective potential $V(Q, T)$ then has a maximum at $Q=0$ and a minimum at $Q=\pi f/2$. The form of the effective potential is reduced to Eq. (\ref{PNGBpot}).  

Let us now see in detail the dynamics of the N=2 PNGB field of SSB. In this work, we consider only small displacements ($Q\ll M_p$) from the top of the N=2 PNGB potential
at $Q=0$. Then the N=2 PNGB effective potential (\ref{PNGBpot}) becomes
\beq
\label{ApproxPoten}
V(Q)=2M^4 -\frac{1}{2}m^2_qQ^2.
\eeq
Here the last term denotes that the field $Q$ near the top of the N=2 PNGB potential corresponds to  the tachyonic term with the negative mass squared $V''(0)=-m^2_q$.

For the approximate potential (\ref{ApproxPoten}), the equation of motion for the N=2 PNGB field $Q(t,\vec{x})$ looks as follows:
\beq
\label{ApproxEQ}
\ddot{Q}-\nabla^2{Q}-m^2_qQ=0~(Q\ll M_p).
\eeq
Here $\nabla$ denotes the spatial derivative with respect to $\vec{x}$. The N=2 PNGB mode functions
$Q_{k}$ thus satisfy
\beq
\ddot{Q}_{k}+(k^2-m^2_q)Q_{k}=0,
\label{ApproxEQ1}
\eeq
where $k\equiv|\vec{k}|$. At early times $t < t_c$ ($t_c$ is a time for the temperature $T$ of the Universe to reach a critical temperature $T_c$ or a late time phase transition to occur), when the temperature $T$ of the Universe is larger than $T_c$,
that is, $T >T_c\sim M$, the $Z_2$ symmetry of the
N=2 PNGB field is completely restored so that the vacuum expectation
value of the field is $\left<Q\right>= 0$. At late time $t=t_c$, when the temperature $T$ is $T=T_c\sim M$, the N=2 PNGB field then becomes massless. When $m_q$ is zero, the mode functions to (\ref{ApproxEQ1}) are given as:
\beq
Q_{k}(t)=\frac{1}{\sqrt{2k}}e^{-ik(t-t_c)}.
\label{Modfun}
\eeq
We take that the mode functions describing quantum fluctuations in the symmetric phase $Q=0$, at the moment close to $t=t_c$ are the same as the one for Eq. (\ref{Modfun}).
At late times $t>t_c$, when the temperature $T$ decreases below $T_c\sim M$, namely $T<T_c\sim  M$, the temperature T of the Universe turns on the tachyonic term, $-\frac{1}{2}m^2_qQ^2$, corresponding to the negative mass squared $-m^2_q$. Consequently, in this quench approximation, all modes $k<m_q$ grow exponentially as follows:
\beq
Q_{k}(t)=\frac{1}{\sqrt{2k}}e^{(t-t_c)\sqrt{m^2_q-k^2}}~(t> t_c).
\label{Quench}
\eeq
Using the Heisenberg picture, the quantum field of the N=2 PNGB is given by
\beq
Q(t, \vec{x})=\int\frac{d^3\vec{k}}{(2\pi)^{3/2}}\left[{a}_{\vec{k}}{Q_k}e^{i\vec{k}\cdot\vec{x}}   + a^{\dag}_{\vec{k}}{Q^*_k}e^{-i\vec{k}\cdot\vec{x}}\right].
\label{PNGBsol}
 \eeq
At $t>t_c$, the modes with $k<m_q$ grow exponentially, and their dispersion is given by \cite{Felder:2001rt} as follows:
\beqa
\label{Disper}
\left<Q^2\right>&=&\left<0|Q (t,\vec{x}) Q(t,\vec{x})|0\right>, \no
&\equiv&\int^{m_q}_0\frac{dk^3}{(2\pi)^3}|Q_k(t)|^2, \no
 &=& \frac{e^{2m_q(t-t_c)}\left[2m_q(t-t_c)-1\right]+1}{16\pi^2(t-t_c)^2}.
\eeqa
At $t=t_c$, the initial dispersion of all growing fluctuations with $k<m_q$ becomes
\beqa
\label{Inidis}
\left<Q^2\right>&=&\left<0|Q (t_c,\vec{x}) Q(t_c,\vec{x})|0\right>, \no
 &=& \int^{m_q}_0\frac{dk^2}{8\pi^2}=\frac{m^2_q}{8\pi^2}.
\eeqa

At $t>t_c$, to characterize the average amplitude of quantum fluctuations of the N=2 PNGB field $Q (t,\vec{x})$ in vacuum state, we use the equal-time correlation function
$\left<0|Q (t,\vec{x}) Q (t,\vec{y})|0\right>$. For a set mode functions $Q_k(t)$ the correlation function is given by
\beq
\label{Correl}
\left<0|Q (t,\vec{x}) Q (t,\vec{y})|0\right>=\int\frac{k^2dk}{2\pi^2}|Q_k(t)|^2\frac{\sin{k|\vec{x}-\vec{y}|}}{k|\vec{x}-\vec{y}|}.
\eeq
By using the relation between the two-point correlation function (\ref{Correl}) and the fluctuations of the averaged field \cite{Mukhanov:2005rt},
\beq
\label{CorrTofluct}
\left<0|Q (t,\vec{x}) Q (t,\vec{y})|0\right>=\int\frac{dk}{k}\delta Q^2_k(t)\frac{\sin{k|\vec{x}-\vec{y}|}}{k|\vec{x}-\vec{y}|},
\eeq
we can infer the average amplitude $\delta Q_k(t)$ of all growing quantum fluctuations for the averaged field with $k<m_q$ in  (\ref{Disper}):
\beq
\delta Q_k(t)=\frac{k}{2\pi}e^{(t-t_c)\sqrt{m^2_q-k^2}}~(t\ge t_c).
\label{Corin}
\eeq
At $t=t_c$ in (\ref{Corin}), the average initial amplitude $\delta Q_k$ of the averaged field becomes
\beq
\label{Amp}
\delta Q_k=\frac{k}{2\pi}.
\eeq


\section{N=2 PNGB QUINTESSENCE AND COSMIC ACCELERATION}
In this section, we show that the late time cosmic acceleration can be driven by the N=2 PNGB quintessence of SSB. Here we will focus on the dynamics of the N=2 PNGB quintessence of SSB and the  fast-roll cosmic acceleration within the self-consistent Hartree approximation \cite{Kinosita:1954rt}.

We now turn to the FRW cosmology which is based on the assumption that the Universe is isotropic and homogeneous on large scales. The dynamics of the flat FRW universe is governed by the Friedman equation:
\beq
H^2=\left(\frac{\dot{a}}{a}\right)^2=\frac{1}{3M^2_p}(\rho_m+\rho_Q),
\label{Hubble}
\eeq
where $a(t)$ is the scale factor with cosmic time $t$, $\rho_m$ and $\rho_Q$ are respectively the energy density of matter and the N=2 PNGB quintessence field $Q$. First of all, we want to calculate the energy density $\rho_Q$ and the pressure $P_Q$ of the N=2 PNGB quintessence field $Q$. To do it, we will follow Ref. \cite{Kinosita:1954rt} and break up the N=2 PNGB quintessence field $Q$ into its expectation value $q(t)$ and fluctuations $\phi$ about that value $q(t)$:
\beq
\label{Homoeq1}
Q(t, \vec{x})=q(t)+\phi(t, \vec{x}), 
\eeq
where
$q(t)$ is $q(t)=Tr[\rho(t)Q(t, \vec{x})]\equiv\left<Q(t, \vec{x})\right>$ and $\rho(t)$ density matrix.
Then the dynamics of the homogeneous component $q(t)$ of the N=2 PNGB quintessence field $Q$ is determined by
\beq
\label{Homoeq}
\ddot{q}+3H\dot{q}-m_q^2q=0~\left(q\ll M_p\right),
\eeq
where $m_q=2(\frac{M^2}{M_p})$.

Hereafter, we will focus on the dynamics of the fast-roll cosmic acceleration in quintessence-dominated era.
The dynamics of the fast-roll cosmic acceleration is related to the following three conditions: first condition (a) is that the equation of state of the N=2 PNGB field should satisfy $P_Q\simeq-\rho_Q$. second condition (b) is that tachyonic instability should be triggered by the presence of tachyonic mass term such as $V''(0)=-m^2_q$ near the top of the potential. third condition (c) is that the quintessence field should satisfy the fast-roll condition $|m^2_q|=O(H^2)$. The quintessence field then starts to roll down from the top of its effective potential toward the potential minimum.

We expect that the N=2 PNGB quintessence field $Q$ acts as the cosmological constant in matter-dominated era and then want to calculate concretely the energy density $\rho_Q$ and the pressure $P_Q$ of the N=2 PNGB quintessence field $Q$. To do this, we consider approximately the simplest possibility, that is, the dynamics of the N=2 PNGB quintessence field of SSB in Minkowski spacetime (neglecting the expansion of the Universe). Then the dynamics of the homogeneous component $q(t)$ of the N=2 PNGB quintessence field $Q$ is determined by
\beq
\label{Homoeq1}
\ddot{q}-m_q^2q=0~\left(q\ll M_p\right)
\eeq
and the solution is
\beq
\label{Homosol2}
q(t)=q_c^{m_q(t-t_c)},~q_c =q(t_c )~\left(t_c\le{t}<t_0\right),
\eeq
where $t_0$ is the onset of the late time cosmic acceleration.
In this case the energy density $\rho_Q$ and the pressure $P_Q$ of the N=2 PNGB quintessence field $Q$
are given by
\beqa
\rho_Q&=&\frac{1}{2}\dot{q}^2+\frac{1}{2}\left<\dot{\phi}^2\right> +\frac{1}{2}\left<(\nabla{\phi})^2\right>+\left<V(Q)\right>, \no
P_Q&=&\frac{1}{2}\dot{q}^2+\frac{1}{2}\left<\dot{\phi}^2\right>-\frac{1}{6}\left<(\nabla{\phi})^2\right>-\left<V(Q)\right>, \no
\eeqa
where the expectation value of the potential energy density $V(Q)$ and the kinetic energy density of the homogeneous field $q(t)$ are respectively given by
\beqa
\label{quntumE1}
\left<V(Q)\right>&=&2M^4 -\frac{1}{2}m^2_qq^2-\frac{1}{2}m^2_q\left<{\phi^2}\right>, \no
\frac{1}{2}\dot{q}^2&=&\frac{1}{2}m^2_qq^2\sim \frac{m^4_q}{8\pi^2}.
\eeqa
Here $\dot{q}=m_qq$ from (\ref{Homosol2}) and the initial field displacement $q$ is $q\sim{q_c}\sim\frac{m_q}{2\pi}$ when $k\sim{m_q}$ in (\ref{Amp}), thus $\frac{1}{2}\dot{q}^2=\frac{1}{2}m^2_qq^2\sim \frac{m^4_q}{8\pi^2}\ll2M^4~(\frac{m_q}{M}\sim 10^{-30})$.
We see that  $V(Q)$ is given by the approximate potential (\ref{ApproxPoten}). In the approximate case it is straightforward to get solution to the Klein-Gordon equation (\ref{ApproxEQ}).  As stated above, solution to the equation is given by the quantum field (\ref{PNGBsol}). Let us use the quantum field to estimate the energy density $\rho_{Q}$ of the N=2 PNGB quintessence field $Q$. The main contribution to the growing quantum fluctuations $\phi$ of the N=2 PNGB quintessence field $Q$ is given by fluctuations $\phi$ with $k\ll m_q$. In the limiting case ($k\ll m_q$), we calculate the mode functions of quantum fluctuations $\phi$ of the field $Q$:
\beq
\phi_{k}(t)=\frac{1}{\sqrt{2k}}e^{m_q(t-t_c)}~(t\ge t_c).
\label{Quench}
\eeq
At $t=t_0$, setting $m_q(t_0-t_c)\sim 1$, the initial dispersion of all growing fluctuations $\phi$ with $k\ll m_q$ is
\beqa
\label{Disper}
\left<\phi^2\right>&=&\int^{m_q}_0\frac{dk^3}{(2\pi)^3}\frac{e^{2m_q\Delta t}}{2k}, \no
&=& \frac{m^2_q}{8\pi^2}e^{2m_q\Delta t}, \no
 &\sim& \frac{m^2_q}{8\pi^2}.
\eeqa
where $\Delta{t}=t_0-t_c$. Substituting the initial dispersion relation (\ref{Disper}) into $\frac{1}{2}m^2_q\left<{\phi^2}\right>$ in (\ref{quntumE1}), we obtain
\beq
\label{Poten2}
\frac{1}{2}m^2_q\left<{\phi^2}\right>\sim\frac{m^4_q}{16\pi^2}.
\eeq
The initial expectation value of the gradient energy density of the fluctuations $\phi$ is
\beqa
\label{Graden}
\frac{1}{2}\left<(\nabla{\phi})^2\right>&=&\frac{m^4_q}{32\pi^2}e^{2m_q\Delta t}, \no
 &\sim&\frac{m^4_q}{32\pi^2},
\eeqa
and the initial expectation value of the kinetic energy density of the fluctuations $\phi$ is given by
\beq
\label{Kinen}
\frac{1}{2}\left<\dot{\phi}^2\right>\sim \frac{m^4_q}{16\pi^2}.
\eeq
From the ratio of the PNGB quintessence field mass to the explicit symmetry breaking energy scale, $\frac{m_q}{M}\sim 10^{-30}$, we find that each of Eq. (\ref{Poten2}), Eq. (\ref{Graden}) and Eq. (\ref{Kinen}) is much smaller than $2M^4$, that is, $\frac{1}{2}m^2_q\left<{\phi}^2\right>$,
$\frac{1}{2}\left<\dot{\phi}^2\right>$, and $\frac{1}{2}\left<(\nabla{\phi})^2\right>\ll 2M^4$. Thus it follows that equation of state of the N=2 PNGB quintessence field is
$P_Q\simeq-\rho_Q$, that is, $w_Q=\frac{P_Q}{\rho_Q}\simeq-1$. Hence first condition (a) is satisfied.

As mentioned in section II, at late times, as the Universe expands, when the temperature $T$ decreases below $M$ (i.e., when $T< M$), then at $t=t_c$, the temperature $T$ of the Universe switch on the tachyonic
mass term $-\frac{1}{2}m^2_qQ^2$ near the top of  the N=2 PNGB potential. Consequently, tachyonic instability in this model is triggered by the presence of tachyonic term
such as $V''(0)=-m^2_q$. thus (b) is also met.

We finally check the fast-roll condition (c). The first term $2M^4$ in Eq. (\ref{quntumE1}) is much greater than $\frac{m_q^4 }{16\pi^2}$, thus we get $\frac{1}{2}m^2_q\left<\phi^2\right>\ll2M^4$. From Eq.
(\ref{Graden}) and (\ref{Kinen}), we also obtains that $\frac{1}{2}\left<\dot{\phi}^2\right>$, $\frac{1}{2}\left<(\nabla{\phi})^2\right>\ll2M^4$, and
hence $\rho_Q\simeq2M^4$. In the N=2 PNGB quintessence-dominated era, the constant $\rho_Q$ makes the Hubble parameter in the Friedman equation (\ref{Hubble}) remain almost unvarying. Substituting $\rho_Q\simeq2M^4$ into the Friedman equation (\ref{Hubble}), neglecting $\rho_m$ and combining $m_q= 2 (\frac{M^2}{M_{p}})$, we obtain that $m^2_q\simeq{6H^2}$. Therefore the fast-roll condition $|m^2_q|=O(H^2)$ is also satisfied.

Before estimating the number of e-folds during the epoch of the late time cosmic acceleration, let us first check for a moment the matter-dominated epoch.
In matter-dominated epoch $\left(t_c\le{t}<t_{eq}\right)$, where $t_{eq}$ denotes dark energy-matter equality time,
the scale factor is given as $a(t) \propto{t}^{\frac{2}{3}}$. At $t\simeq t_c$, matter energy density is given as $\rho_{m}>\rho_Q\simeq 2M^4$. At late times, as the Universe expands, the matter energy density with $w=1/3$ falls off as $\rho_{m}\propto{a^{-3}}$, and the N=2 PNGB quintessence energy density with $w_Q\simeq-1$ is constant as $\rho_Q\propto{a^0}$. As a result, at $t>t_{eq}$, it is reversed to $\rho_Q >\rho_{m}$. 

We have showed that the fast-roll cosmic acceleration triggered by the N=2 PNGB quintessence of SSB is responsible for an epoch of the late time cosmic acceleration. Let us now calculate the total exponential  expansion of our universe during the epoch of the late time cosmic acceleration.
At the current epoch $t=t_0$,
the energy density of the N=2 PNGB quintessence $\rho_{Q}(t_0)$ is
order of magnitude $\rho_{crit}(t_0)$, i.e.,
 $\rho_{Q}(t_0)\simeq\rho_{crit}(t_0)\simeq2M^4$. In the N=2 PNGB quintessence-dominated era, because the energy density of the N=2 PNGB
quintessence $\rho_Q(t_0)$ and the Hubble parameter $H$ in the Friedman equation (\ref{Hubble}) remain almost unvarying, consequently the solution to Eq. (\ref{Homoeq}) is given as
\beq
q(t)=q_0e^{\alpha(y){H(t-t_0)}}~(t\ge{t_0}),~q_0 = q(t_0),
\label{Homosol}
\eeq
and
\beq
\alpha(y)=\sqrt{y+\frac{9}{4}}-\frac{3}{2},~ y=\frac{m^2_q}{H^2}.
\label{Alpha}
\eeq

As mentioned earlier, classically, the N=2 PNGB field could settle down initially on the top of the effective
potential at $Q = 0$. But we see that quantum fluctuations $\phi$ of the N=2 PNGB quintessence field prevent us from taking $q_c=0$, since $q_c$ can't be taken smaller than an average initial amplitude $\delta\phi_{k}(t_c)$ of quantum fluctuations with momenta $k<m_q$. We set the initial field displacement $q_c$ to $q_c\sim\delta\phi_{k}(t_c)$ $\sim\frac{m_q}{10}$ \cite{Linde:2001rt}. In addition the current field displacement $q_0\sim{q_c}\sim\frac{m_q}{10}$; here we used $m_q\Delta{t}\sim 1$ and $q_0=q_c e^{m_q(t_0 -t_c)}$.
In Eq. (\ref{Homosol}) we have field velocity $\dot{q}=\alpha{H}q$, and find that $\dot{q_0}\ll q_0$, and initial conditions $q_0$, $\dot{q_0}$ thus satisfy $\dot{q_0}\ll{q_0}\ll{M_p}$.
Eq. (\ref{Homosol}) and Eq. (\ref{Alpha}) allow us to evaluate the number of e-folds for the total exponential expansion of our universe during the era of the late time cosmic acceleration:
\beq
e^{H(t-t_0)}\sim\left(\frac{10q}{m_q}\right)^{\frac{1}{\alpha(y)}}.
\label{Efold}
\eeq
We see that $\left<V(Q)\right>\simeq V(q)=2M^4-\frac{1}{2}m^2_q{q}^2$. Here we neglected the $\left<\phi^2\right> + higher~order ~terms$ of the inhomogeneous component $\phi$ of  the N=2 PNGB quintessence field $Q$. For the approximate potential $V(q)$, the late time cosmic acceleration is valid until $q(t_1)=10^{-3}{M_{p}}$ because $q(t_1)$ is much less than ${M_p}/\sqrt2$: here we used the potential condition for the accelerated expansion, $\frac{1}{2}M^2_p(\frac{V'}{V})^2\ll 1$. This is a 
good approximation. Then we have $m^2_q\simeq{6H^2}$ and hence we find that $y=6$, $\frac{1}{\alpha(6)}=0.73$ and $ m_q\simeq\sqrt{6}H\thicksim2\times10^{-60}M_p$. From (\ref{Efold}), we obtain
\beq
\label{Total}
e^{H(t_1-t_0)}\thicksim\left(10^{57}\right)^{0.73}\thicksim10^{42}\thicksim e^{96}.
\eeq
Thus, Eq. (\ref{Total}) shows that the fast-roll cosmic acceleration in the N=2 PNGB quintessence is responsible for up to $96$ e-folds of the total exponential expansion of our universe during the epoch of the late time cosmic acceleration.

\section{DISCUSSIONS AND CONCLUSIONS}

In this paper, we have explored the possibility that in the N=2 PNGB quintessence model initial field configuration to respect $Z_2$ symmetry could result from the finite-temperature effects on the effective potential. However, this is not a realistic possibility. We expected that the N=2 PNGB field rolled quickly down to the minimum of the effective potential at $Q=0$ before the late time phase transition provided $c(T)$ is sufficiently large and the N=2 PNGB field mass $m_q$ is sufficiently large as well. But in radiation-dominated and matter-dominated era it is expected that the N=2 PNGB field mass $m_q$ is less than the Hubble parameter $H$. Thus it is unlikely that the N=2 PNGB field rolled quickly down to the minimum of the effective potential at $Q=0$ before the late time phase transition.

Let us briefly refer to a more realistic possibility, hybrid quintessence. 
If we introduce an auxiliary field which interacts the N=2 PNGB field, the N=2 PNGB field classically can then be trapped at $Q=0$ by such a mechanism as hybrid inflation \cite{Linde:1994rt}. Hence, a more detailed analysis of our model and associated problems need to be addressed in the context of hybrid quintessence: for example hybrid dark energy model in which an axion monodromy interacts with another axion quintessence with SSB. We are going to address hybrid dark energy model in the following future work. 

However, in the N=2 PNGB quintessence model, one still might worry about the domain wall problem. Unlike conventional phase transitions that the transitions occur at a high energy scale, that is, before decoupling of CMB, in the N=2 PNGB model, the late time phase transition occurs at most the $M=O(1\rm{meV})$ energy scale, and at redshift $z\sim10$ after decoupling of CMB. Such low energy transitions can lead to phenomenologically acceptable density inhomogeneities in domain walls with minimal variations in the microwave anisotropy \cite{Hill:1989rt}. Also in our PNGB model, the late time cosmic acceleration can be quite long lasting since the mass of the N=2 PNGB quintessence field is extremely small as $m_q \approx 10^{-33} \rm{eV}$: indeed, in the N=2 PNGB quintessence, the late time cosmic acceleration is responsible for up to $96$ e-folds of the total exponential expansion of our universe during the epoch. Thus we now don't need to worry about the problem since domain walls is created after $96$ e-folds. Even if $Z_2$ symmetry is spontaneously broken down after the late time cosmic acceleration there may be the Lazarides-Shafi mechanism Ref. \cite{Laza:1982rt,Chatterjee:2020rt} to avoid problems with domain wall. A detailed study of the domain wall problem in this model is beyond the scope of this article.

We have proposed an idea that dark energy is originated from the N=2 PNGB quintessence of spontaneous symmetry breaking. We have investigated the dynamics of the N=2 PNGB quintessence of SSB and the fast-roll cosmic acceleration which happens when the quintessence field starts to roll down fast from the top of its effective potential toward the potential minimum.

We have found that in the N=2 PNGB quintessence model, the fast-roll cosmic acceleration associated with SSB allows the accelerating expansion of the Universe to occur. It has been shown that the fast-roll cosmic acceleration, in the N=2 PNGB quintessence, is responsible for up to $96$ e-folds of the total exponential expansion of the Universe during the epoch of the current cosmic acceleration. Indeed, in the N=2 PNGB model, this epoch is sustained quite long since the mass of the N=2 PNGB quintessence field is extremely small as $m_q \approx 10^{-33} \rm{eV}$.

In N=1 PNGB quintessence models, the late time cosmic acceleration could be only achieved by fine-tuning the N=1 PNGB field to make the quintessence settle down near the top of its effective potential for a long time to simulate the cosmological constant.
But in the N=2 PNGB quintessence model we have shown that SSB applied to the N=2 PNGB quintessence slightly alleviates not only the initial condition problems but also the imperfect late time cosmic acceleration of N=1 PNGB quintessence models.

The most important consequence in this work is the observation that the first stage of the late time cosmic acceleration occurs not due to the slow-rolling down of the quintessence field but due to the fast-rolling down from the top of the quintessence field associated with spontaneous symmetry breaking. Although we have used the idea and observation on the N=2 PNGB quintessence for the dark energy problem, it seems to be applicable to the problems on the inflationary initial condition.

\section*{ ACKNOWLEDGMENTS }
This work was supported by the National Research Foundation of Korea (NRF) grants funded by the Korea government(MSIT) (No. NRF-2020R1F1A1068410).

\end{document}